\documentclass[aip,apl,reprint,showpacs,10pt,amsmath,amssymb,floatfix]{revtex4-1}
\pdfoutput=1

\usepackage{amsmath,amssymb,graphicx} 

\usepackage{upgreek}

\begin{document}
\title{Multiple double-metal bias-free terahertz emitters}

\author{D. McBryde}
\email{duncan.mcbryde@soton.ac.uk}
\author{P. Gow}
\author{S. A. Berry}
\author{M. E. Barnes}
\author{A. Aghajani}
\author{V. Apostolopoulos}
\affiliation{School of Physics and Astronomy, University of Southampton, Southampton, SO17 1BJ, United Kingdom}

\begin{abstract}
We demonstrate multiplexed terahertz emitters that exhibits 2 THz bandwidth that do not require an external bias.
The emitters operate under uniform illumination eliminating the need for a micro-lens array and are fabricated with periodic Au and Pb structures on GaAs.
Terahertz emission originates from the lateral photo-Dember effect and from the different Schottky barrier heights of the chosen metal pair.
We characterize the emitters and determine that most terahertz emission at 300 K is due to band-bending due to the Schottky barrier of the metal.
\end{abstract}


\maketitle


\begin{figure}[b]
\centering
\includegraphics[width=2in]{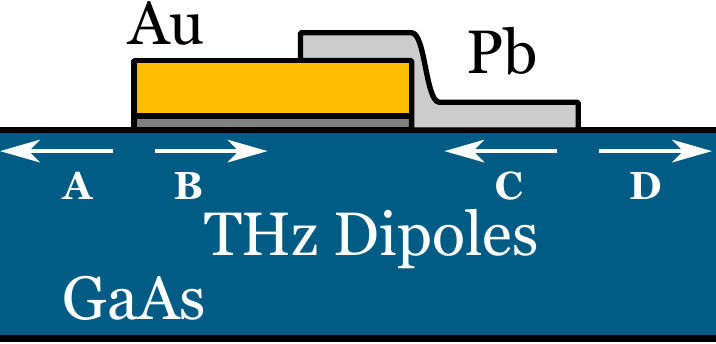}
\caption{
A diagram showing the emission mechanism of multiple-metal emitters due to the lateral photo-Dember effect.
Carrier diffusion creates radiative dipoles near the metal boundaries, shown as arrows.
Each set of dipoles created on the boundary are labelled A, B, C and D.
The difference in the reflectivity between the two metals that quench dipoles B and C causes net terahertz emission to be observed.
}
\label{fig:lpd}
\end{figure}

Lateral photo-Dember (LPD) emitters have been demonstrated as robust terahertz emitters requiring no voltage bias to operate \cite{Barnes2012, McBryde2014, Klatt2010}. 
In an LPD emitter a semiconductor surface is partially masked with a deposited metal layer.
An ultra-fast laser, with above band-gap energy, is focused half on the metallic mask and half on the semiconductor surface, thus, creating an asymmetrical distribution of photo-generated carriers near the metal-semiconductor interface that is free to diffuse.
The hypothesis for LPD emission was that the asymmetric carrier concentration was responsible for a net diffusion current \cite{Klatt2010}.
In \cite{Barnes2012, Barnes2013} we have shown that terahertz emission co-linear with the optical pump due to an anisotropic carrier concentration is not possible because diffusion creates net zero current. 
Therefore diffusion currents cannot be the sole mechanism for radiation. We have completed the hypothesis of Klatt et al. \cite{Klatt2010} by attributing the mechanism of radiation of LPD emitters to diffusion currents and to dipole quenching with a metal mask \cite{Barnes2012, Barnes2013}.
The carriers that diffuse under the metal mask cannot radiate due to the proximity to the metallic surface in relation to the radiating wavelength. The rest of the carriers that diffuse away from the metal are free to radiate co-linearly with the pump beam in the same geometry as a photo-conductive antenna.
As LPD emitters depend on lateral currents, surface field effects in the semiconductor do not contribute to terahertz emission.

\begin{figure}[b]
\centering
\includegraphics[width=3.37in]{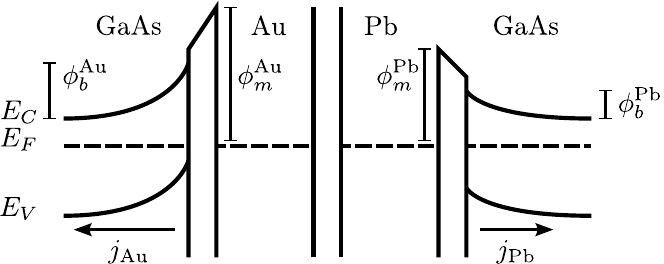}
\caption{
A band diagram of the repeatable terahertz emitters.
Both gold and lead create a Schottky barrier near the interface.
Two metals have different work functions, $\phi_m$, creating different barrier heights $\phi_b$.
Band bending occurs near the metals creating currents $j_\text{Au}$ and $j_\text{Pb}$.
If $j_\text{Au} \neq j_\text{Pb}$ terahertz emission due to a net current is generated.
}
\label{fig:schottky}
\end{figure}

It is useful to multiplex the LPD emitters to increase their performance, however, periodic single-metal ridges under uniform illumination can not produce net terahertz emission in the direction of detection \cite{Barnes2013}, each set of dipoles created either side of the ridges is suppressed equally producing a quadrupole emission pattern.
This can be alleviated by illuminating alternating metal edges \cite{Gow2013}.
Furthermore, multiple LPD emitters have previously been successfully demonstrated with uniform illumination by fabricating a periodic gold wedge pattern \cite{Klatt2010, Klatt2011}.  
The multiple emitters in \cite{Klatt2010} were created because it was thought that the metallic wedges would generate terahertz emission due to the generation of periodic anisotropic carrier distributions.
Taking into account the dipole quenching mechanism \cite{Barnes2012, Barnes2013} we now attribute the terahertz emission observed in \cite{Klatt2010, Klatt2011} to the reduced quenching strength caused by the reduction in metal height on one side of the mask. 
In this work, we propose to achieve the required asymmetry for co-linear THz emission by applying a low skin depth second metal on one side of each repeat of the metal mask, reducing the quenching strength from one side of the metal mask by reducing the reflectivity of the metal surface \cite{Drexhage1970}.
In doing so we eliminate the requirement for a micro-lens array used previously \cite{Gow2013}, allowing for the emitters to operate under uniform illumination.
The reduced quenching strength allows net terahertz emission to be formed in a dipole pattern under the low skin depth metal.
Ideally the metal is opaque to the optical pump beam and is transmissive to terahertz emission so quenching is reduced to a minimum.
An opaque dielectric may be used in place of an additional metal at the expense of increased difficulty in emitter fabrication.
This geometry is shown in Fig.\ \ref{fig:lpd} where gold completely quenches the dipole formed underneath whereas lead is deposited with a thickness below its skin depth for THz frequencies and therefore not inhibiting the terahertz dipole to radiate.
The dipoles are shown in Fig.\ \ref{fig:lpd} and are labelled A, B, C and D.
Dipoles A and D oppose each other and only produce quadrupole emission.
Dipole B is fully quenched by the gold mask.
Dipole C is not fully quenched as the Pb layer has less reflectivity in the terahertz region.
Under uniform illumination terahertz emission is observable co-linear with the surface normal due to the difference in quenching strength from the two metals.

Band bending near the surface of a semiconductor can cause terahertz emission \cite{Zhang1990} and has been shown to influence LPD emitters in \cite{McBryde2014} and to modify the polarity of terahertz pulses under intense optical fluence \cite{Shi2006, McBryde2014}.
Therefore, in the repeatable structures that we show here we used metal pairs that would also create terahertz emission due to different Schottky barrier potentials.
By selecting metals with different skin depths and work functions both the LPD effect and Schottky barriers can contribute to terahertz emission.
The concept is shown in Fig.\ \ref{fig:schottky} where by depositing metals with different work functions, $\phi_m$, we create different barrier heights on each side of the double-metal emitters.
The band bending causes a transient lateral current to form near the metal boundary emitting terahertz emission.
If the barrier heights $\phi_b$ are different, net lateral current will occur proportional to the difference in barrier height under uniform illumination.
If the barrier heights are equal, as the case would be for a single metal, no net current would be present under uniform illumination.
If the barrier heights are not equal the relative strength of dipoles A and D in Fig.\ \ref{fig:lpd} differ, producing net terahertz emission co-linear with the surface.
In the multiple LPD emitters that we present the requirement for any focusing lens for the optical pump is removed, allowing the emitters to act as a drop-in replacement for photo-conductive emitters.
These emitters are simple to fabricate, do not require an electrical bias or a silicon lens to out-couple the THz emission.
We expect that these emitters are more durable than conventional photo-conductive emitters due to no risk of electrical breakdown.


\begin{figure}
\centering
\includegraphics[width=2.5in]{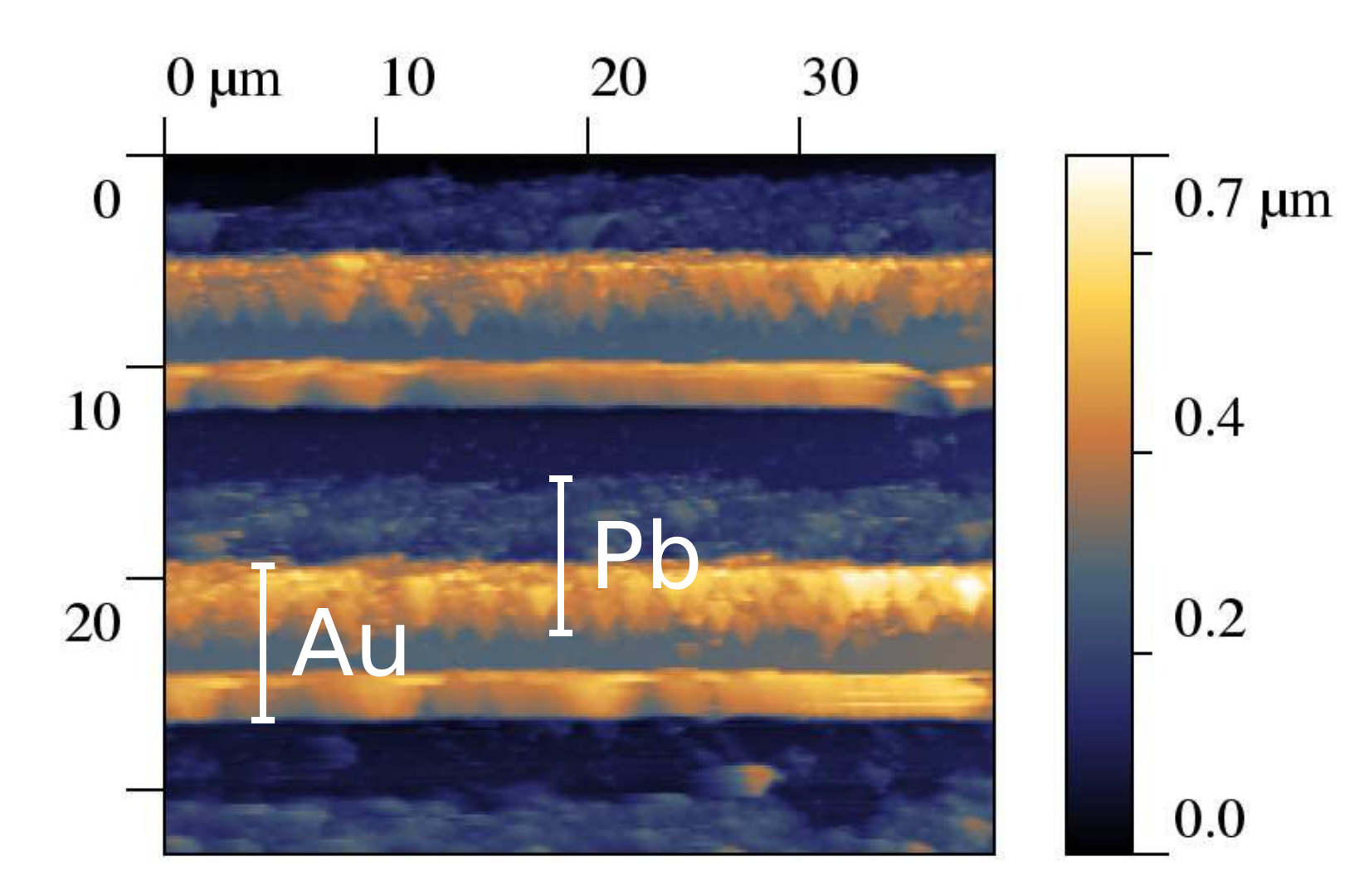}
\caption{
An atomic force microscope image of the repeatable emitters fabricated with gold and lead.
The lead layer has oxidized with the air, forming a rough surface.
}
\label{fig:afm}
\end{figure}

Our target were emitters that utilize both diffusion and Schottky generated currents on semi-insulating (SI) GaAs. Therefore, we chose the metals in order for both to reflect the optical pump but only one of them to exhibit metallic behaviour in the THz region.
We also chose the two metals in order to exhibit a high difference in work functions.
Au was chosen as the high reflecting metal as it has a skin depth of 80~nm at 1 THz \cite{Ordal1985} and was deposited at a thickness of 170~nm, Cr was used as an adhesive layer but at a thickness (3 nm) that the Schottky potential is still dominated by Au.
The second chosen metal was Pb which possesses a high skin depth in the terahertz region, 238~nm at 1 THz \cite{Ordal1988}.
Pb layers with thickness below 24~nm are transparent at 1 THz \cite{Singh2008}, which we verified in our THz-TDS experiment.
Because we wanted Pb to be opaque to the 800 nm pump laser we chose for the fabrication to deposit a thickness of 100~nm.
Each metal strip was 4~$\upmu$m wide with a 15~$\upmu$m period and a 2~$\upmu$m overlap between the Au and Pb, as shown in Fig. \ref{fig:afm}.
Due to Bardeen surface states the barrier height is mostly independent of the work function of the metal but is dependant on surface quality \cite{Bardeen1947} due to oxidizing layers formed on the semiconductor surface.
The barrier height, $\phi_b$ of gold bonded with GaAs is approximately 0.9~eV \cite{Mead1964,Spitzer1963}, and lead is 0.8~eV at 300~K \cite{Waldrop1984}.
This potential difference can create a net lateral carrier current.
We also fabricated the same emitters with a 150~nm SiO$_2$ insulating layer to eliminate band bending from the metal \cite{Nishimura2008} in order to characterize the relative strength of the LPD and Schottky effects.
The multiple double-metal emitters were tested in a standard THz-TDS experiment using a Ti:Sapphire pump laser; centre wavelength 800~nm, 80~MHz repetition rate with a 70~fs pulse length.
A Menlo Tera-8 photo-conductive antenna was used as a receiver with a focusing Si lens.
The double-metal emitters were mounted without any focusing Si lenses.
The emitters were illuminated with the pump laser with an beam diameter of 1~mm, illuminating 67 emitters.
This spot size allows the emitter to act as a planar wave rather than a point source emitter for lens-free operation.
An objective lens and a neutral density filter allowed the intensity and spot size to be adjusted, allowing the optical fluence on the multiple double-metal emitters to be varied.


\begin{figure}[htp]
\centering
\includegraphics[width=3.37in]{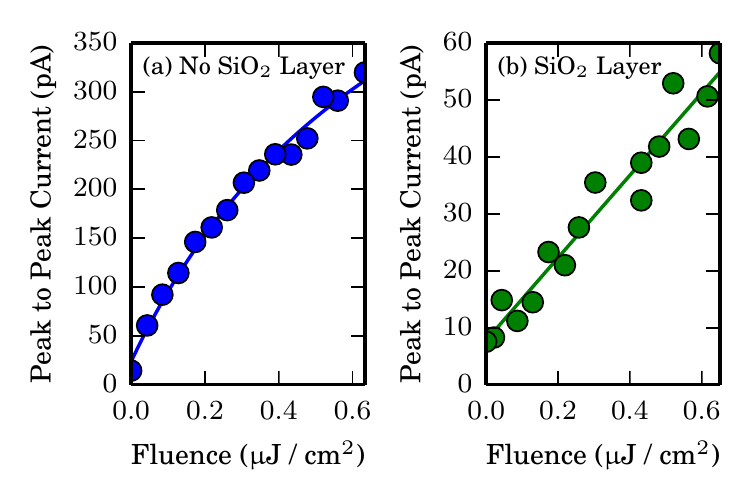}
\caption{
(a) shows the fluence dependence of Au/Pb without an SiO$_2$ layer and (b) with an insulating SiO$_2$ layer.
The optical spot radius is held at 430 $\upmu$m.
The saturation fluence, $F_\text{sat}$, (a) is fitted from equation \ref{eqn:fit} and determined to be 0.82~$\upmu$J cm$^{-2}$, (b) shows no saturation.
}
\label{fig:fluence}
\end{figure}

\begin{figure}[htp]
\centering
\includegraphics[width=3.37in]{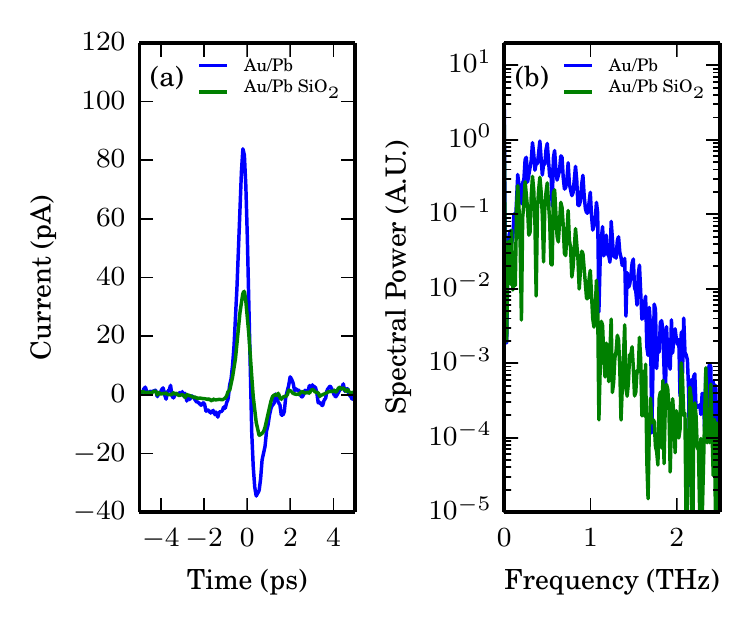}
\caption{
(a) time domain scans for the Au/Pb emitters insulated with SiO$_2$ and Au/Pb uninsulated, (b) the corresponding power spectrum.
Measurements taken at room temperature.
}
\label{fig:comparison}
\end{figure}

We characterized the Au/Pb emitters with fluence to test for saturation effects.  We used a saturation fit of the form

\begin{equation}
E(F) = A \frac{F}{F + F_\text{sat}},
\label{eqn:fit}
\end{equation}

\noindent where $F$ is the optical fluence and $F_\text{sat}$ is the optical saturation fluence, $A$ is a coefficient that describes efficiency.
More terahertz emission was observed for emitters directly fabricated on SI-GaAs compared with emitters fabricated on SI-GaAs with an insulating SiO$_2$ layer.
$F_\text{sat}$ was determined to be 0.82~$\upmu$J cm$^{-2}$ for metal directly fabricated on SI-GaAs, whereas no saturation was observed for double-metal structures fabricated with an insulating layer.
The value of $A$ from the non-SiO$_2$ emitters was 20 times greater compared with the SiO$_2$ fabricated emitter, showing that most of the terahertz emission from the double-metal emitters is from the Schottky interface. 
The cause of saturation within the directly bonded double-metal emitters is likely to be due to the charge accumulation within the depletion region. The drift current of the electrons from the metal to the semiconductor is reduced for higher fluences, at high carrier concentrations the barrier height decreases due to the higher carrier density and higher Fermi energy in the semiconductor  \cite{Shi2006, McBryde2014}.

\begin{figure}[htp]
\centering
\includegraphics[width=3.37in]{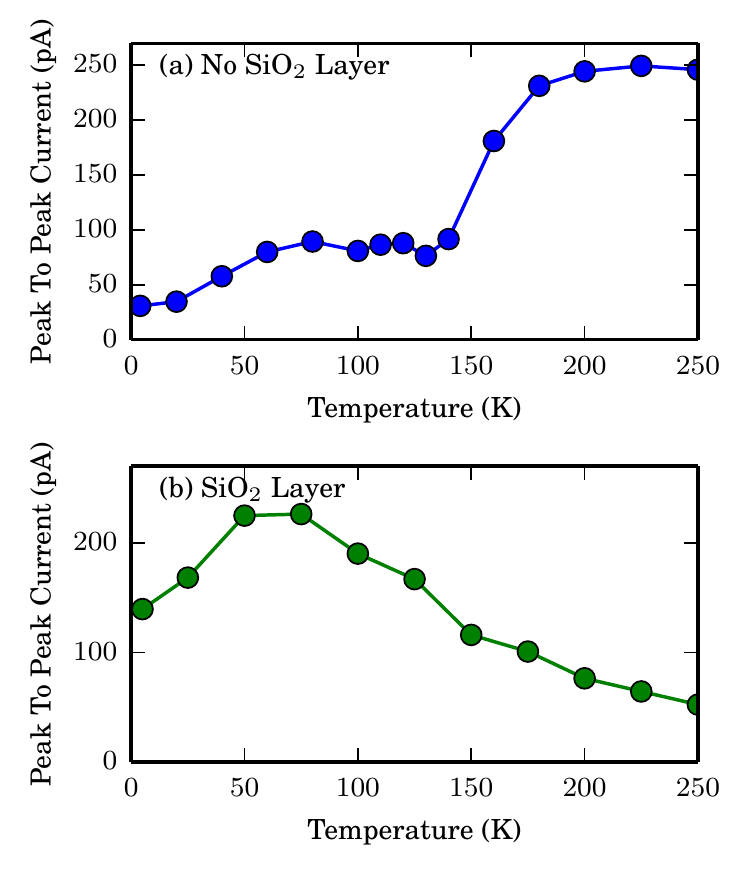}
\caption{
Temperature dependence of Au/Pb and Au/Pb insulated LPD emitters.
(a) shows the temperature dependence for the emitters without an SiO$_2$ layer and  (b) shows the temperature dependence for the emitters with an insulating SiO$_2$ layer. 
}
\label{fig:temperature}
\end{figure}

The time domain scans and  the corresponding power spectrum for the Au/Pb insulated with SiO$_2$ and Au/Pb uninsulated are shown in Fig.\ \ref{fig:comparison} for optical pump power of 180~mW and spot diameter of 1~mm at room temperature. 
The results show that the most powerful emitters are Au/Pb fabricated directly of SI-GaAs, the Au/Pb insulated emitters as mentioned above, is weaker working only on the difference of quenching efficiency of Au and Pb stripes. 
Double-metal emitters demonstrate a dependence on the optical pump polarisation similar to the one encountered in \cite{Huggard1998, McBryde2014}, all the measurements presented here are made with a polarisation perpendicular to the metal edges.
Multiple double-metal emitters have similar signal to noise performance compared with large gap SI-GaAs photoconductive antennas measured in a terahertz time-domain spectroscopy (THz-TDS) experiment.
The double-metal emitters produce emission of 2 THz bandwidth with 33 dB of dynamic range as shown in Fig.\ \ref{fig:comparison} similar to a SI-GaAs large gap photoconductive emitter biased at 20 V with a 140 $\upmu$m electrode gap.


We measured terahertz emission power with temperature for both Au/Pb SiO$_2$ and non-SiO$_2$ fabricated emitters.
Both emitters were mounted in a helium flow cryostat.
Fig.\ \ref{fig:temperature}~(a) and (b) show the peak-to-peak THz signal with temperature for the non-SiO$_2$ and SiO$_2$ double-metal emitters respectively.
The alignment of the emitters changed with the mounting in the cryostat, so the detected emission viewed in Fig.\ \ref{fig:temperature} (a) and (b) is not strictly comparable.
The terahertz output power increases with decreasing temperature for the insulated SiO$_2$ emitter, with an optimum emission temperature at 50-80~K, shown in Fig.\ \ref{fig:temperature} (b).
This  increase in output power is attributed to the increased electron mobility at low temperatures \cite{Wolfe1970}, as fewer free carriers are present the free electron path length is increased. 
This implies carrier mobility is the main source for THz emission when Schottky barrier emission is removed.
Both emitters exhibit a decrease in THz output below 50~K which is in accordance with the reduction in electron mobility in GaAs \cite{Wolfe1970}.
In the case of the uninsulated emitters there is a contribution from the Schottky barriers formed at the Au/GaAs and Pb/GaAs interfaces as well as emission due to diffusion. These emitters show reduced performance when the temperature is reduced below 200~K (Fig.\ \ref{fig:temperature} (a) ). This follows  the temperature relationship of the Au/GaAs Schottky barrier height \cite{Hudait2001}.
Between 50~K and 100~K the THz emission levels out and experiences a small rise in power.
This coincides with the temperature region in which carrier mobility reaches its peak values and Schottky barrier height approaches its minimum values \cite{Wolfe1970, Hudait2001}.


We have demonstrated robust, simple to fabricate terahertz emitters based on both the LPD effect and the Schottky barrier effect.
The main source for emission at room temperature is due to band bending from the Schottky barrier. We also show that the LPD effect plays a role in the terahertz emission that becomes more apparent at low  temperatures of 50-80 K in accordance with increased electron mobility. 
We demonstrate effective terahertz emission but there is scope for improvement by choosing different metal pair to optimize the amount of Schottky currents. 
However, because the Schottky barrier heights are dependent on material and metal deposition it is difficult to predict the strength of the emission without fabricating the emitters.
The amount of LPD emission can also be optimized further by choosing different metals and thicknesses to enhance the quenching difference between the chosen metal pairs.
Due to their lack of voltage bias requirements double-metal terahertz emitters allow for passive terahertz generation and increased robustness and may suit applications where wire-bonding would be difficult.

This work was supported by the EPSRC, grants EP/G05536X/1 and EP/J007676/1.

Copyright 2014 American Institute of Physics. This article may be downloaded for personal use only. Any other use requires prior permission of the author and the American Institute of Physics.
The following article has been accepted by Applied Physics Letters. After it is published, it will be found at \url{http://scitation.aip.org/content/aip/journal/apl}.

%

\end{document}